\documentclass[preprint2]{aastex}

\slugcomment{version June 17, 2003}

\newcommand{\etal}{\textit{et~al.}}

\newcommand{\fn}[1]{\footnote{\scriptsize{#1}}}
\shorttitle{Centaurs}
\shortauthors{Tiscareno and Malhotra}

\begin{document}

\title{The Dynamics of Known Centaurs\fn{In press at \textit{The Astronomical Journal}}}
\author{Matthew S. Tiscareno, Renu Malhotra}
\affil{Lunar and Planetary Laboratory, University of Arizona, Tucson, AZ 85721;\\ matthewt@lpl.arizona.edu, renu@lpl.arizona.edu}

%\textbf{ABSTRACT}
\begin{abstract}

We have numerically investigated the long term dynamical behavior of known Centaurs.  This class of objects is thought to constitute the transitional population between the Kuiper Belt and the Jupiter-family comets~(JFCs).  In our study, we find that over their dynamical lifetimes, these objects diffuse into the JFCs and other sinks, and also make excursions into the Scattered Disk, but (not~surprisingly) do not diffuse into the parameter space representing the main Kuiper Belt.  These Centaurs spend most of their dynamical lifetimes in orbits of eccentricity 0.2-to-0.6 and perihelion distance 12-to-30 AU.  Their orbital evolution is characterized by frequent close encounters with the giant planets.  Most of these Centaurs will escape from the solar system (or enter the Oort Cloud), while a fraction will enter the JFC population and a few percent will impact a giant planet. Their median dynamical lifetime is 9~Myr, although there is a wide dispersion in lifetimes, ranging from less than 1 Myr to more than 100 Myr.  We find the dynamical evolution of this sample of Centaurs to be less orderly than the planet-to-planet ``hand-off'' described in previous investigations.  We discuss the implications of our study for the spatial distribution of the Centaurs as a whole.

\vspace{1cm}

\end{abstract}

\section{Introduction \label{Intro}}

In the past decade, there has been a rapid increase in the number of discoveries of a transitional population of minor planets in the outer solar system called the Centaurs.  The first object in this population, Chiron, was discovered in 1977 \citep{Kowal}, and several dozen are now known, most discovered within the last 5 years.  These objects are characterized by highly chaotic orbits with perihelion lying between Jupiter's orbit and Neptune's orbit.  Their dynamical lifetimes are much shorter than the age of the solar system, thus they must have a source in a more stable reservoir elsewhere in the outer solar system.  The prevailing view is that Centaurs are objects that have escaped from the trans-Neptunian Kuiper Belt and represent the dynamical population intermediate between the relatively stable Kuiper Belt source and the short-lived short period Jupiter family comets (JFCs).  This picture is based upon a number of theoretical investigations that have explored the Kuiper~Belt--JFC connection by means of numerical simulations \citep{Duncan87, Duncan88, Holman93, LD93, LD97, DL97, Morbidelli}.

While the dynamics of Kuiper Belt objects (KBOs) has been, and continues to be, a hot topic of investigation, relatively little attention has been given to the dynamics of the Centaur population itself.  The study by \citet{LD97}, hereafter LD97, provides perhaps the best previous understanding of the dynamics of this class of objects.  These authors used a numerical model to trace the evolution of objects escaping from the Kuiper Belt into the Jupiter family comets, including their transition as Centaurs.  In common with most other past studies, LD97 assumed that the source population in the Kuiper Belt is dynamically ``cold,'' i.~e., that their orbits are nearly circular and of low inclination.  Observations over the last decade, however, have increasingly indicated that the Kuiper Belt is not dynamically cold; rather, Kuiper Belt objects (KBOs) have a surprisingly broad distribution of orbital eccentricities and inclinations.  Furthermore, it consists of several distinct dynamical populations: the resonant KBOs, the non-resonant main belt (or classical) KBOs and the Scattered Disk objects (see \citet{Malhotra00} for a recent review).  Thus, the KB--JFC connection certainly needs to be revisited with dynamical models that take account of the relatively ``excited'' and complex Kuiper Belt structure.  In the present work, we have the more limited goal of studying the dynamics of the population intermediate between KBOs and JFCs, namely, the Centaurs.  We use the known Centaur population as our starting point.  We study their long term dynamics with the goal of understanding their dynamical history and eventual fate, their connection to the short period comets and to the presumed Kuiper Belt source(s), and the present overall distribution of Centaurs.  

Our numerical model is described in Section~\ref{Model}.  In Section~\ref{Results}, we describe our results, including details of the diversity of orbital histories of the known Centaurs, quantitative estimates of their dynamical lifetimes, and statistics of close encounters with the outer planets.  In Section~\ref{Implications}, we discuss the implications of our results for the distribution of the entire Centaur population.  We summarize our conclusions in Section~\ref{Conclusions}.

\begin{figure}[!t]
\includegraphics[angle=90,width=8cm,keepaspectratio=true,viewport=7cm 0cm 16cm 21cm]{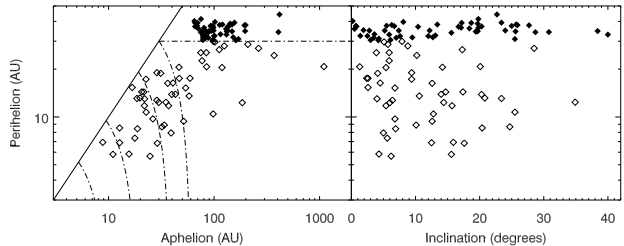}
\caption{Orbital parameters of known Centaurs and Scattered Disk Objects (SDOs).  The 53~Centaurs, which were used as the initial parameters for our simulation, are shown as open diamonds, while the SDOs are shown as solid diamonds.  In the left-hand figure, the horizontal dash-dot line (which divides the two populations) represents perihelia equal to that of Neptune; while the curved dash-dot lines represent lines of constant semimajor axis, with values corresponding to Saturn, Uranus, and Neptune.  Data from the Minor Planet Center website, as of 2002~May~1.  \label{InitParams}}
\end{figure}

\section{Model \label{Model}}

We obtained the orbital parameters of the known Centaurs from the Minor Planet Center~(MPC)\fn{http://cfa-www.harvard.edu/iau/lists/Centaurs.html}.  As of 2002~May~1, the MPC listed 110~Centaurs and Scattered-Disk objects (SDOs) in a single table.  The orbital distribution of this sample is shown in Fig.~\ref{InitParams}.  Of these 110 objects, we chose a subset of 53 objects which have perihelion distance interior to Neptune's orbit.  We define these as Centaurs, as distinct from SDOs which have perihelion distance exterior to Neptune's orbit.  (Although this choice based on perihelion distance alone is reasonable as a rough dynamical division between Centaurs and SDOs, a more rigorous definition of the distinction between the more transient Centaurs and the long-lived SDOs would be useful for future detailed studies. In section~\ref{Results} we suggest a modest improvement of this definition, based on our results.)  The epochs of the Centaurs, as given by the MPC, range from 2002~September~20 to 2002~May~6.  For the purpose of an efficient numerical study, we calculated the initial conditions of these objects at the common epoch of 2000~January~1, assuming unperturbed Keplerian motion on their present orbits.  

In our sample, the median eccentricity is~0.483 and the median inclination is~$9.4^\circ$.  For those Centaurs that cross only Neptune, the Tisserand parameter\fn{The Tisserand parameter of a particle, with respect to a particular planet, remains relatively unchanged through encounters with that planet.  $T = a_P/a + 2 \sqrt{ (1-e^2) a/a_P } \cos i$, where $a_P$ is the planet's semimajor axis, and $a$, $e$, and $i$ are the test~particle's semimajor axis, eccentricity, and inclination, respectively.} with respect to Neptune~($T$) is significantly below 3 (ranging from 2.46 to 2.92). This reflects the dynamically hot nature of this sample.  For comparison, LD97's theoretical model started with a much more dynamically cold sample: 18~of their 21~initial particles have $T$ between 3.01 and~3.08, and none have $T$ lower than~2.82.  (We must compare the initial conditions in this manner because LD97 report the initial conditions of their model only as KBOs that are about to encounter Neptune.)

We modeled the Centaurs as massless test particles, and we followed their orbital evolution for 100~Myr under the perturbations of the outer four planets, Jupiter through Neptune.  The mutual perturbations of the planets were calculated self-consistently in our modeling.  We stopped following a particle once it reached $r < 2.5$~AU, the dynamical region of visible comets; and we consider a particle to have been ejected from the solar system upon reaching $r > 20,000$~AU.  For the numerical orbit integrations, we used ``Swift-Skeel,'' a mixed-variable symplectic $N$-body integrator with the capability to handle close encounters between test particles and planets \citep{Duncan98, Wisdom}.  Our integrations were performed with a step size of 0.1~Earth years, and we recorded the position and velocity of each particle every 20,000~years. (As we will see, the total integration time of 100~Myr is adequate for our study, as the mean dynamical lifetime of our sample of Centaurs is found to be only 9 Myr.)  It is important to note that, in light of the strongly chaotic orbital dynamics of these objects, individual particle histories in our results should not be viewed as the determined orbital history of any single object; rather, they should be viewed statistically, in the context of their overall time-weighted distribution.

In the analysis that follows, we often plot on a single figure the accumulated record of orbital parameters in our 100-Myr integration (with the 20,000-year time resolution, as stated above).  Such figures illustrate the time-weighted cumulative orbital parameter distributions of our sample of Centaurs over their dynamical lifetimes.  In section~\ref{Implications}, we discuss the relationship between these distributions obtained from our simulation and those of the entire Centaur population, considering various observational selection effects and modeling assumptions.

\section{Results \label{Results}}

\begin{figure}[!t]
\includegraphics[angle=90,width=16cm,keepaspectratio=true]{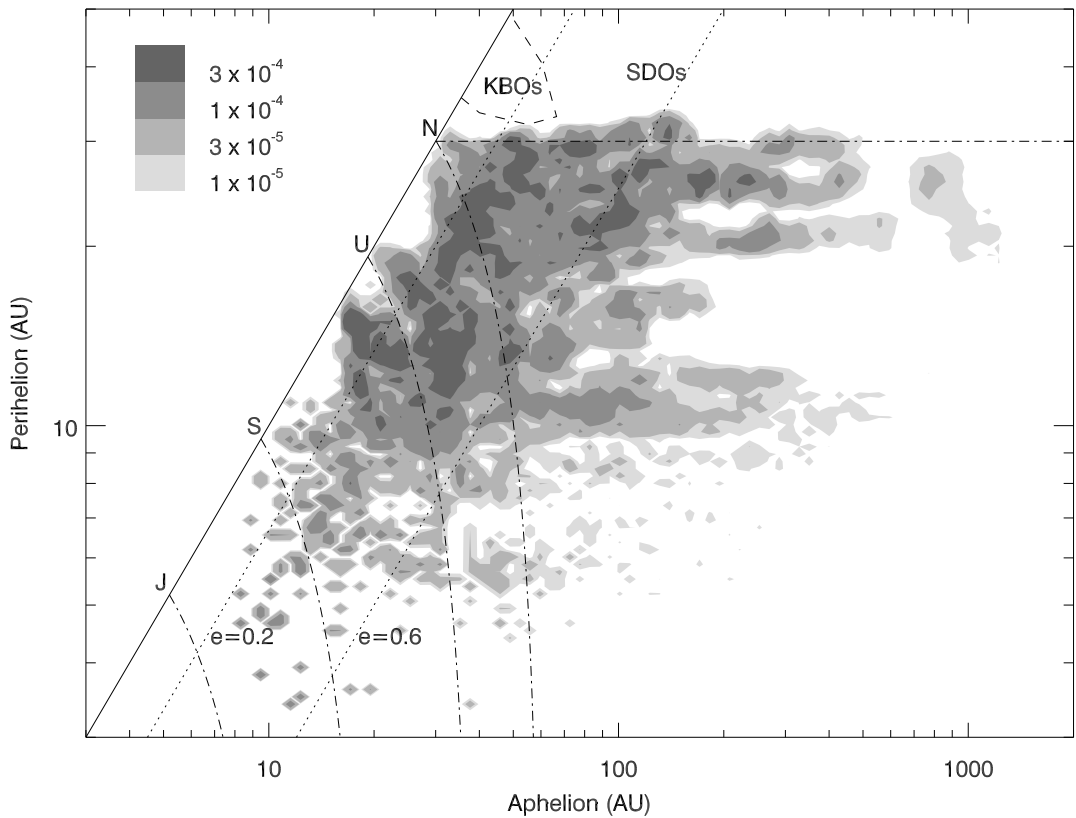}
\caption{The time-weighted distribution of the known Centaurs obtained in our integration.  The shading, as shown in the legend to the upper left, is the fraction of particles per AU$^2$.  The solid diagonal line corresponds to orbits of zero eccentricity; thus, anything to the left of that line is unphysical.  Similarly, the dotted diagonal lines represent lines of constant eccentricity.  The curved dot-dash lines represent lines of constant semimajor axis, with values corresponding to the four giant planets.  The horizontal dot-dash line represents orbits of perihelion distance equal to that of Neptune.  The region contained within the dashed lines at the top-center is the appproximate location of the classical and resonant KBO populations (some of the latter actually extend to smaller perihelion distances, though this is not reflected in the figure). The remaining region of perihelion distances close to but exceeding that of Neptune defines the main zone of SDOs. \label{Contour}}
\end{figure}

A general overview of the dynamical evolution of the observed sample of Centaurs, as determined in our integrations, is shown in Fig.~\ref{Contour}.  Here we have combined the records of all our test-particle Centaurs (at 20,000-year intervals over the 100-Myr total integration time) and have obtained their time-averaged distribution as a function of the perihelion ($q$) and aphelion ($Q$) distances.  The figure shows contours of this density in gray-scale, as the fraction of objects per square AU in the ($q,Q$) plane.  (Note that this is not the ``surface density'' in physical space, but in dynamical parameter space.)  The density variations reflect the relative lengths of time that our sample of Centaurs spend in various regions of this parameter space.  

A prominent feature in Fig.~\ref{Contour} is the strong drop-off in density beyond perihelion distance $\sim33$~AU; there is very little diffusion of these Centaurs to $q\gtrsim33$~AU.  This outer boundary is likely related to the boundary of the chaotic zone of overlapping first-order mean motion resonances of Neptune \citep{Malhotra96}.  Although $q=30$ AU is presently adopted as the boundary between Centaurs and SDOs, the excursions of the known Centaurs to $q>30$ AU over their dynamical lifetimes indicate some, or possibly significant, overlap with the phase space of the Scattered Disk, depending upon the precise definition of the latter.  Alternatively the apparently strong diffusion barrier identified here suggests a natural dynamical division between Centaurs and the Scattered Disk at $q\approx33$ AU.  This is consistent with previous studies of the dynamics of SDOs \citep{DL97}. In future studies, detailed mapping of the diffusion rates near this boundary could define the boundary more precisely.

Furthermore, we note that there is essentially no penetration of these objects into the main region of resonant and classical KBOs (the approximate boundaries of the latter are indicated by the dashed lines in the upper-center of the figure). This is not entirely surprising: dynamically long-lived regions, by their nature, are characterized by relatively non-porous boundaries; particles are able to slowly ``leak'' out of such a heavily populated, dynamically stable region into less stable regions, but their diffusion rate back into the more stable region is very small, as they are instead dispersed through the solar system.  Thus, this apparent lack of visitation of our Centaurs into the main domain of KBOs does not necessarily preclude the latter as a possible source of these Centaurs.  For

\onecolumn

\noindent the same reasons, the diffusion barrier between Centaurs and SDOs discussed in the previous paragraph does not preclude the SDOs as a source of Centaurs.

The right-hand side of Fig.~\ref{Contour} is dominated by horizontal (constant perihelion) features.  Three of these correspond to perihelion values near the locations of Saturn, Uranus, and Neptune, respectively; these most likely represent the main paths of ejection, as particles suffer close encounters with the planets near their own perihelia, and their aphelia are gradually pumped up to large values.  An analogous feature due to Jupiter is also discernible, but is less prominent in this figure due to low particle densities and short timescales of ejection there.  This type of evolution has been noted in previous studies of the formation of the Oort Cloud \citep{Duncan87, Fernandez97}.  Another horizontal feature, in the perihelion range from 23~to 27~AU, does not appear to be linked with any single planet or a favored path of ejection.  This 

\begin{figure}[!t]
\center{\includegraphics[angle=90,width=12.5cm,keepaspectratio=true]{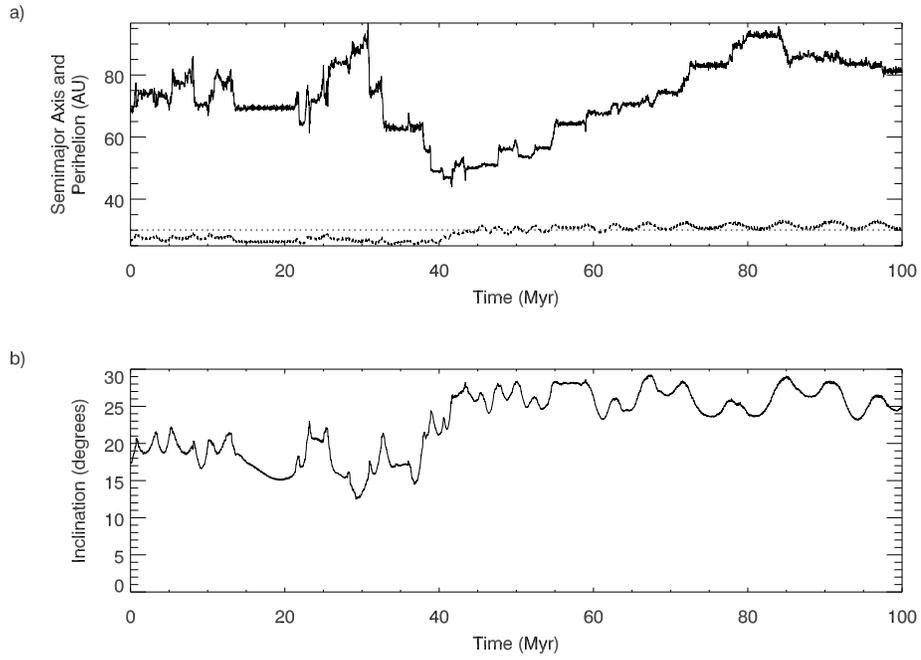}}
\caption{Individual history of one of the 53~particles in our model.  The starting parameters correspond to those of 2002~CY$_{224}$.  \label{P1}}
\end{figure}
\begin{figure}[!t]
\center{\includegraphics[angle=90,width=12.5cm,keepaspectratio=true]{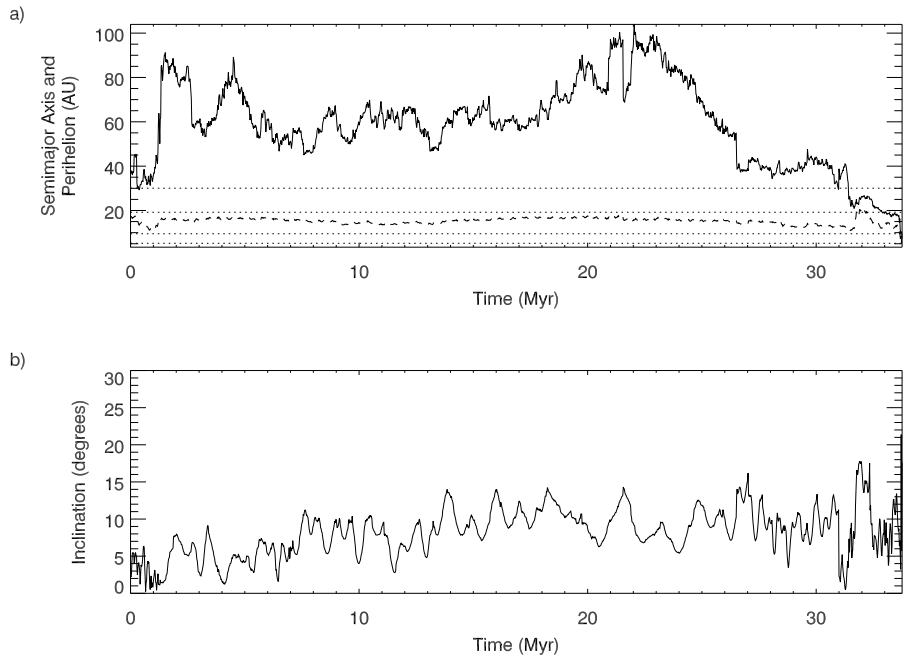}}
\caption{Individual history of one of the 53~particles in our model.  The starting parameters correspond to those of 2002~CR$_{46}$.  \label{P2}}
\end{figure}
\twocolumn

\noindent region is characterized by higher inclinations than average (see Section~\ref{Radial} and Fig.~\ref{Icontour}), and may be a pocket of relative stability.  

The figure shows that these Centaurs spend most of their time at moderate eccentricities (0.2~to~0.6) with perihelia from 12~to 30~AU.  Their distribution is fairly uniform across this area, with a density contrast of only a factor of $\sim30$.  This indicates that the objects diffuse freely throughout this region.

In some contrast with the high and nearly uniform density at modest eccentricities, there is a relatively low density at small eccentricities.  This suggests relatively high chaotic diffusion rates at $e \lesssim 0.2$ compared to the rest of the Centaur phase space.  We note that the results of LD97 similarly show a relative low density at low eccentricities (see their Fig.~6).  Several stability studies indicate that low eccentricity, low inclination orbits are unstable in the Jupiter-to-Neptune region on timescales of $10^4$--$10^7$ yr \citep{Grazier, Lecar}.  Although there is moderate visitation of parts of the low eccentricity region in Fig.~\ref{Contour}, we find that this is mainly due to the particles that begin with low eccentricities (see Fig.~\ref{InitParams}) but are perturbed into higher-eccentricity ($e \gtrsim 0.2$) orbits after just a few Myr.  The detection of several objects in this region is therefore somewhat surprising; we attribute it to an observational bias in favor of detecting low eccentricity objects (see Section~\ref{fair}).

\subsection{Dynamical History \label{DynHist}}

The individual histories of two of the 53 particles in our study are shown in Figs.~\ref{P1} and~\ref{P2}.  These two cases exemplify the range of behaviors seen in our simulations.  

The particle in Fig.~\ref{P1} exhibits ``resonance hopping''; that is, it jumps amongst various mean-motion resonances with Neptune, but does not remain in a single resonance for any long period of time.  Beginning with an inclined, eccentric, Neptune-crossing orbit, this particle spends at least 28 separate discrete periods of time in at least 18 different resonances (characterized by a librating value of the semimajor axis).  Its longest single stay in a resonance is only about 8~Myr.  Yet, despite its apparent lack of stability, the particle survived to the end of the integration (100~Myr).  Like many of the particles in our simulation that move from resonance to resonance, this particle spends most (97.3\%) of its time with an inclination higher than $15^\circ$.  It remains Neptune-crossing for most of its history, and its perihelion never exceeds 34~AU.

In contrast, the particle in Fig.~\ref{P2} exhibits no discernible ``resonance hopping'' or ``resonance sticking'' of any kind (although we cannot rule out ``resonance hopping'' with residence times shorter than $\sim 100,000$~yr).  Its semimajor axis fluctuates apparently randomly, beginning at 38.3~AU, to over 100~AU, before it is injected into the inner Solar System.  This particle spends almost all of its history in orbits that cross both Uranus and Neptune, while maintaining a moderate inclination (it spends only 1.3\% of its time with an inclination higher than $15^\circ$, although some particles exhibiting this type of extended non-resonant behavior do have higher inclinations).  The particle survives for 34~Myr under this arrangement, before being injected into the inner solar system.

The other particles in our sample exhibit some combination of these two behaviors, either ``resonance hopping'' or eschewing resonances altogether.  Several particles range quite a bit more widely in semimajor axis than, and most do not survive quite as long as, the two examples shown.  None stay in any single resonance for longer than several~Myr.  This behavior is similar to that found by \citet{Dones} in their integrations of the orbits of 2~short-period comets and 4~Centaurs; however, it is in marked contrast to the dynamics of SDOs, where ``resonance sticking'' is stronger, and particles are likely to remain in a single resonance over Gyr timescales \citep{DL97}.  We interpret this to indicate that stable resonance islands take up a much smaller fraction of Centaur phase space than is the case for SDOs.  

\begin{figure}[!t]
\includegraphics[angle=90,width=8cm,keepaspectratio=true,viewport=9cm 12cm 18cm 23cm]{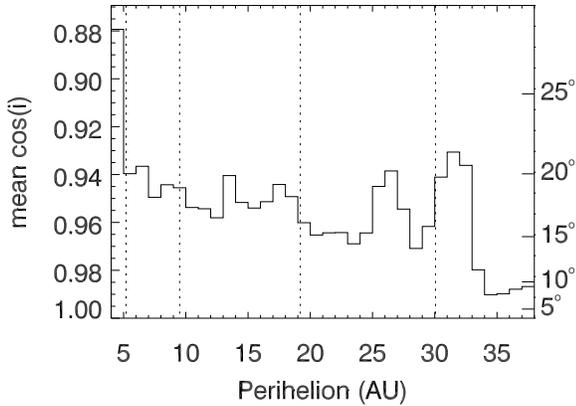}
\caption{The time-weighted inclinations vs.~perihelia of the Centaurs in our simulation.  The locations of the giant planets are shown as dashed vertical lines. \label{qi}}
\end{figure}

LD97 describe a process in which KBOs are ``handed off'' from the gravitational influence of one planet to another, Neptune to Uranus to Saturn to Jupiter, steadily inward towards the JFC population.  One piece of evidence cited to support this is that, for all particles crossing a particular giant planet, LD97's results show a median Tisserand parameter~$T$ with respect to that planet between 2.6~and~2.8 (A value of $T \lesssim 3$ indicates that the particle's dynamics is dominated by that planet.)  Another piece of evidence cited by LD97 is that the inclinations, when plotted versus perihelion, show a dip just exterior to each of the four giant planets (Fig.~4 in LD97); this is expected for $T \sim 3$ which precludes high inclinations.  
Our results are not so simply interpreted.  
In Fig.~\ref{qi} we plot the time-weighted inclination distribution vs.~perihelion found in our integrations; we see no decreases associated with any of the giant planets.  
In our simulation, the median Tisserand parameters for particles crossing Jupiter, Saturn, Uranus, and Neptune (as described above) are 1.89, 2.51, 2.57, and 2.61, respectively.  These values
are systematically and significantly less than those found in LD97's model.
These lower $T$ values make it less likely for particles to remain under the dominant influence of a single planet, much less to be ``handed'' from one planet to another.  ``Hand-off'' behavior probably does occur to some extent amongst these Centaurs, but it does not appear to be the dominant characteristic in their evolution; we find their dynamics to be much less orderly.

\begin{figure}[!t]
\includegraphics[angle=90,width=8cm,keepaspectratio=true,viewport=9cm 12cm 18cm 23cm]{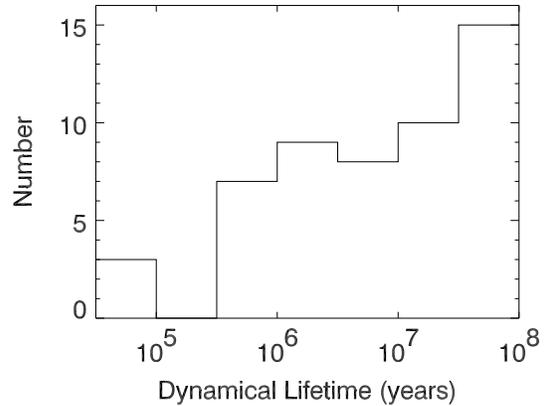}
\caption{Dynamical lifetimes of the 53~test particle Centaurs in this simulation. \label{LifeHist}}
\end{figure}

\subsection{Dynamical Lifetime \label{Lifetimes}}

The median dynamical lifetime of our sample of Centaurs in this simulation is 9~Myr.  This is somewhat longer than the estimate of 1~to 5~Myr given by \citet{Dones}, who studied only 6~objects in the inner (more~chaotic) part of the Centaur region.  The lifetimes do vary widely, with 11 of the 53 particles surviving less than 1~Myr, and another 11 surviving for more than 60~Myr. The distribution of lifetimes is shown in Fig.~\ref{LifeHist}.  Particles are removed from the simulation in two ways:  31~of the 53 particles are ejected from the Solar system (or enter the Oort Cloud), while 15~of the~53 are injected into the inner solar system.  We conclude from this that the former outcome is approximately twice as likely as the latter for these Centaurs.  

During their lifetime, most objects make several transitions amongst dynamical subclasses.  \citet{Levison96} defines the dynamical subclass of JFCs as having a Tisserand parameter with respect to Jupiter, $T$, in the range $2 < T < 3$; while Centaurs, or ``Chiron-type'' comets (after the first such body to be discovered), have $T > 3$ and semimajor axis greater than that of Jupiter.  We tracked the number of times that each object transitions from one class to another.  We find the average length of stay in the ``Chiron-type'' class to be 6.5~Myr, while the same for the JFC class is only 50,000~years.  We also find that our sample of Centaurs spend 98.9\% of their time in the ``Chiron-type'' class, and 0.7\% of their time as JFCs.

We also find transitions into all three of the classes of ``nearly-isotropic'' comets defined by \citet{Levison96}, especially the ``returning'' long-period comets (semimajor axis between 20 and 10,000~AU).  Our sample of Centaurs spend 0.2\% of their time as ``returning nearly-isotropic'' comets.  The inclinations of this subset are indeed nearly isotropic, ranging from $2.2^\circ$ to $173^\circ$, with a mean of $59.6^\circ$.  Their perihelion distances range from 0.037~AU to 6.9~AU, with a mean of 1.9~AU.  

We saw no transitions into the ``Encke-type'' category (aphelia inside Jupiter).  As discussed by LD97 (who also noted this phenomenon), there are several possible explanations for this:  either the mechanism creating Encke-type comets involves effects not included in this model, such as the terrestrial planets or non-gravitational effects (see, for example, \citet{Fernandez02}); or Encke-type comets are so rare that our model does not have sufficient statistics to see them; or Encke-type comets do not originate in the Centaur population.

\subsection{Planetary Encounters \label{Encounters}}

The number of encounters (defined as an approach within a planet's Hill radius) experienced by these 53~test particle Centaurs during their dynamical lifetimes is as follows:  4,743~encounters with Neptune (57\% of the total); 2,053~with Uranus~(25\%); 1,176~with Saturn~(14\%); and 344~with Jupiter~(4\%).  Dividing the total number of encounters by the sum of the lifetimes of the test particles yields an average rate of 1~encounter per particle per 10~Myr.  Two objects came close enough to a planet to impact it, from which we estimate that $(4 \pm 2) \%$ of this sample impacts a planet.  For comparison, LD97 estimate 1.5\% of their population impacts a planet, while \citet{Dones} estimate 1\%.  All three of these numbers, however, suffer from poor statistics. 

Of the 31~particles in our simulation that were ejected from the solar system (or entered the Oort Cloud), 9 were ejected subsequent to an encounter with Jupiter; 11 after encountering Saturn; 2 after encountering Uranus; and 9 after encountering Neptune.  Combining these numbers with the encounter statistics in the previous paragraph, we can estimate that 2.6\% of the encounters of our sample of Centaurs with Jupiter result in ejection; 0.94\% for Saturn; 0.097\% for Uranus; and 0.19\% for Neptune.

\begin{figure}[!t]
\includegraphics[angle=90,width=8cm,keepaspectratio=true,viewport=9cm 12cm 18cm 23cm]{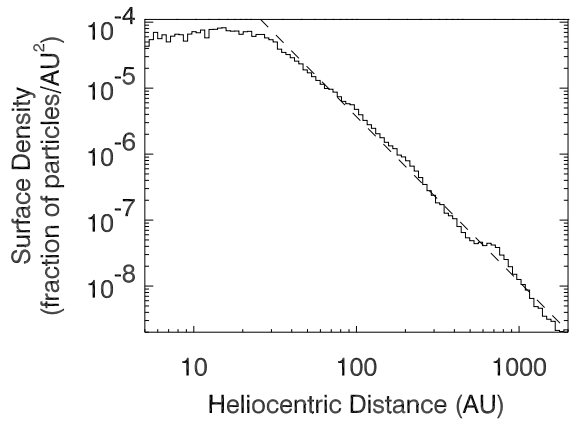}
\caption{The time-weighted radial distribution of the known Centaurs obtained in our simulation.  The surface density is plotted as a function of heliocentric distance in the ecliptic plane. The dashed line shows the best power-law fit for heliocentric distance $r>30$~AU, which corresponds to a power law of $\sim r^{-2.5}$.  \label{Rdistrib}}
\end{figure}

\subsection{Radial and Inclination Distributions \label{Radial}}

The time-weighted radial distribution obtained in our simulation is shown in Fig.~\ref{Rdistrib}, where we plot the surface density as a function of heliocentric distance in the ecliptic plane.  The surface density is nearly constant in the planetary zone, and drops off beyond 30~AU as $r^{-\alpha}$, where $\alpha = 2.5 \pm 0.1$.  This power law is somewhat shallower than that found by LD97 ($\sim r^{-2.8}$).

\begin{figure}[!t]
\includegraphics[angle=90,width=8cm,keepaspectratio=true,viewport=9cm 12cm 18cm 23cm]{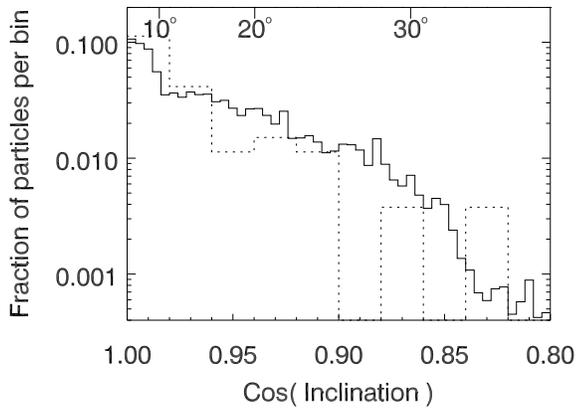}
\caption{The time-weighted distribution of inclinations obtained in our simulation (solid line), and the inclination distribution of the known Centaurs (dotted line).  For ease of comparison, the latter histogram is scaled to a bin size of 0.004.  \label{Idistrib}}
\end{figure}

The time-weighted inclinations of our sample of Centaurs over their dynamical lifetimes are shown in Fig.~\ref{Idistrib} (solid line).  We find that the abundance of inclined orbits declines by a factor of 3 from $0^\circ$ to $10^\circ$.  This is followed by a gentler decrease between $10^\circ$ and $35^\circ$, and a sharper drop-off beyond that.  We also note that the characteristic inclination (defined as the arc-cosine of the mean of $\cos i$) increases with time during our simulation, from $15.5^\circ$ to $19.6^\circ$.  High inclinations are somewhat more abundant in the time-weighted distribution than in the observed sample (dotted line).  A future detailed analysis of observational biases could test how the time-weighted distribution compares with models of inclination-dependent detection bias, such as one proposed by \citet{Brown}. 

\begin{figure}[!t]
\includegraphics[angle=90,width=8cm,keepaspectratio=true,viewport=0cm 0cm 18cm 21cm]{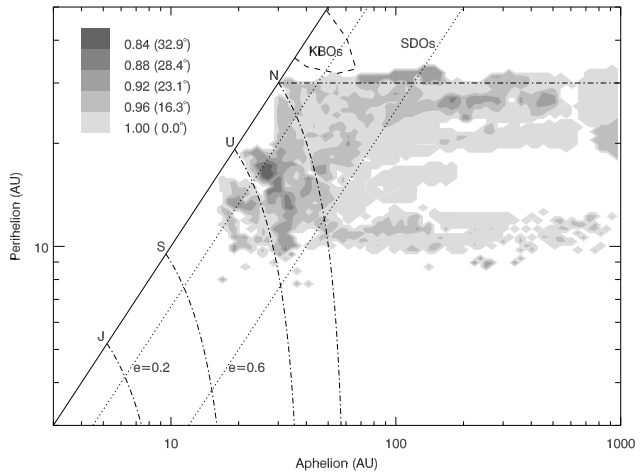}
\caption{The distribution of inclinations in dynamical parameter space.  The shading, as shown in the legend to the upper left, corresponds to the mean of the cosines of the inclinations in each bin.  Only those bins containing more than 5~particles are plotted.  Dashed, dotted, and dot-dash lines are as described for Fig.~\ref{Contour}. \label{Icontour}}
\end{figure}

The spatial distribution of inclinations obtained in our simulation is shown in Fig.~\ref{Icontour}.  In this figure, we plot in gray-scale the time-weighted mean of $\cos i$ in 1 AU-by-1 AU bins in the $(q,Q)$ plane.  Refering back to Fig.~\ref{Contour}, we can see that the region where our sample of Centaurs spends the most time ($e\approx 0.2$--0.6, $q\approx12$--30 AU) has characteristic inclinations, $\cos^{-1}\langle\cos i\rangle$, in excess of $16^\circ$.  High inclinations are also found on the right-hand side of the diagram, among high eccentricity orbits with perihelia near Neptune (overlapping the SDO region) as well as in the high eccentricity feature characterized by perihelia between 23~and 27~AU (see Section~\ref{Results}).  Interestingly, the highest inclinations are found in a region centered on semimajor axes just exterior to that of Uranus and eccentricities up to $\sim0.6$.  Based on our simulation, this is one of the most-visited regions by the known Centaurs over their dynamical lifetimes, as evidenced by the high densities in this region in Fig.~\ref{Contour}.  We can now understand these high densities as being due to the high inclinations which decrease the frequency of planetary encounters and thereby decrease the rate of chaotic diffusion.  The apparent dearth of observed Centaurs here (see Fig.~\ref{InitParams}) can be at least partially understood since the region is characterized by high inclinations, which makes objects less likely to be discovered in ecliptic surveys.  

\section{Implications for the intrinsic Centaur population\label{Implications}}

We have described in the previous section our results on the dynamics of the known sample of Centaurs, based on the distributions of their time-weighted orbital elements over their dynamical lifetimes.  In this section, we consider the implications of these results for the entire Centaur population, most of which remains undetected at present.

The distributions obtained from our simulation, shown in Figs.~\ref{Contour}, \ref{qi}--\ref{Icontour}, would reflect those of the entire Centaur population provided that
\begin{itemize}
\item[i.] the Centaurs' dynamics are nearly ergodic and time-independent; that is, the time-weighted orbital parameter distributions (over the collective dynamical lifetimes of a small random sample of Centaurs) are characteristic of the entire (large) population of Centaurs in steady state between sink and source; and 
\item[ii.] the observed sample is characteristic of the actual Centaur population.
\end{itemize}

For completeness, we should also note that an additional assumption is that the Centaurs' orbital evolution is dominated by the gravitational perturbations of the outer four planets only. The gravitational perturbations not included in our model (e.g., due to the terrestrial planets and other minor planets) are negligible for our purposes, but other effects (such as tidal break-up during close encounters or the effects of outgassing) may also affect the orbital evolution; these are not included in our modeling.  

\subsection{Are Centaur dynamics ergodic?\label{ergodic}}

We do find in our simulation that the orbital evolution of our sample of Centaurs is sufficiently chaotic on timescales much shorter than our 100 Myr integration length, such that, as an ensemble, it is reasonably close to time-invariant. The 100-Myr integration time is long enough to allow a wide range of dynamical behavior to develop.  `Snapshots' of the orbital distribution obtained in our simulation do not change significantly after $\sim5$ Myr have elapsed.  Our simulation has also shown that resonance sticking is not a dominant phenomenon in the dynamics of the known Centaurs, indicating that their dynamics can be described reasonably well as a random walk or diffusion process, i.e. nearly ergodic.  However, we find that there are regions of parameter space where extrapolation of our results to the actual Centaur population is limited by the finite length of time of the simulation.  Specifically, orbits of high eccentricity ($e\gtrsim0.8$) with $q$ in the Saturn-to-Neptune region have dynamical timescales too long for a 100-Myr integration to adequately probe \citep{Wiegert, Malyshkin}.  On the other hand, for orbits of low-to-moderate eccentricities and perihelion distances in the Jupiter-to-Neptune range, the 100 Myr length of our integration is a factor $\sim\!10$~to~10$^4$ longer than the characteristic dynamical lifetimes.

Additional support for ergodicity of this population is obtained by considering a subset of particles that pass through a narrow range of semimajor axis\fn{This test was suggested to us by Hal Levison.}, e.g., $13<a<18$ AU.  The time-weighted orbital distribution of this subset is found to be virtually identical to that of the entire set.  Significant discrepencies are found only on the border with the SDO phase space, near $q \sim 30$~AU and $e \gtrsim 0.8$.  A possible reason for this is that, as noted in the previous paragraph, high-eccentricity regions have slower diffusion rates than any other region of Centaur phase space, and thus have dynamical timescales too long for our 100 Myr simulation to probe adequately. For all other regions of Centaur phase space, the assumption of ergodicity is supported by this test.

\begin{figure}[!t]
\includegraphics[angle=90,width=8cm,keepaspectratio=true,viewport=0cm 12cm 18cm 24cm]{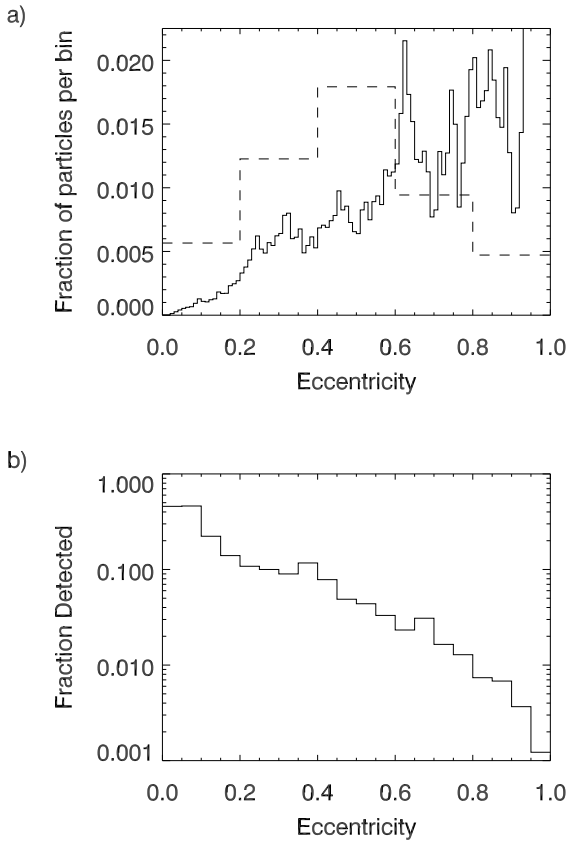}
\caption{(a) Eccentricity distribution of the known Centaurs (dashed) and our time-weighted integration results (solid).  For ease of comparison, both curves are scaled to a bin size of 0.01.  (b)~The fraction of particles detected by the simulated observational survey described in Section~\ref{fair}, as a function of eccentricity. \label{Ebias}}
\end{figure}

\begin{figure}[!t]
\includegraphics[angle=90,width=8cm,keepaspectratio=true]{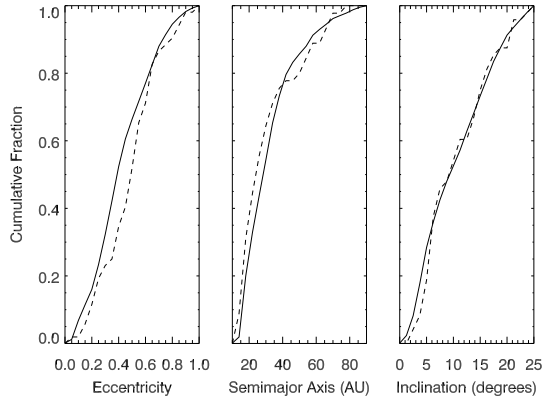}
\caption{Orbital element distributions of the observed Centaur population (dashed), and of a simulated observational sample of our time-weighted integration results (solid).  \label{ObsSim}}
\end{figure}

\subsection{Are the known Centaurs a ``fair sample'' of the intrinsic Centaur population?\label{fair}}

The orbital distribution of the known Centaurs has not been rigorously analyzed for observational biases. Of known concern is the possible under-representation of high inclination orbits, since most known Centaurs have been discovered in ecliptic surveys as part of KBO searches.  \citet{Brown} has evaluated the inclination bias in the KBO population; a generally similar analysis would apply for the Centaurs.

There is also a possible eccentricity bias.  The known sample of Centaurs has a relative paucity of both low and high eccentricity orbits.  In contrast, the time-weighted distribution obtained in our simulation has a heavy proportion of moderate to large eccentricities.  The eccentricity distributions of the observed sample and our time-weighted model population are shown in Fig.~\ref{Ebias}a.

To check whether our time-weighted model population is a possible model for the intrinsic Centaur population, we simulate an observational survey of our time-weighted model population as follows.  We randomly assign to each particle in the model population an absolute magnitude $H$, from the distribution $N(<\!H) \sim 10^{aH}$ with $a=0.7$ \citep{Gladman}.  Then, based on the heliocentric distance of the particle, we calculate its apparent magnitude $m$.  Finally, we consider a particle to be ``observed'' if it has $m < m_{lim}$, and ecliptic latitude $\beta < \beta_{max}$.  For the results described below, we adopted an absolute magnitude range $5\!\leq\!H\!\leq\!10$, a limiting magnitude $m_{lim} = 24$, and a maximum ecliptic latitude $\beta_{max}=5^\circ$.  These are consistent with the range of $H$ for the known Centaurs, and the fact that, roughly speaking, most objects have been detected in ecliptic surveys of limiting magnitude near 24 (e.g., \citet{Millis}).  

Fig.~\ref{Ebias}b shows the ``eccentricity bias'' of the simulated survey applied to our time-weighted model population; it plots the number of detected particles divided by the number of model particles in each eccentricity bin.  It shows that low-eccentricity particles are much more likely to be detected than high-eccentricity ones.  Qualitatively, this result makes sense: the modeled Centaurs span a narrow range in perihelion distance and a wide range of semimajor axis; thus, for a given perihelion distance, objects in low eccentricity orbits remain at small heliocentric distances for relatively longer periods of time than those in higher eccentricity orbits, thus enhancing their detectability.  

In Fig.~\ref{ObsSim} we see that the distributions of orbital elements, $a,e,i$, obtained in the simulated survey (continuous lines) match well those of the observed sample of Centaurs (dashed lines).  This suggests that the time-weighted model population obtained in our simulation is a possible model for the intrinsic Centaur population.  However, it is important to recognize that it may not be a unique model, and that other models of the intrinsic Centaur population may also be compatible with the observed sample.  Only a more detailed bias analysis, which is beyond the scope of the present paper, can establish the confidence limits for such an interpretation.

To summarize:  We conclude from this analysis that our initial conditions (the known Centaurs) are likely biased towards low eccentricities and low inclinations.  However, the system is ergodic in most of Centaur phase space (excluding only the highest eccentricities, $e \gtrsim 0.8$). Therefore, we expect that the time-weighted orbital distribution obtained from our simulation is not extremely sensitive to the biases in initial conditions, and we consider that it provides a reasonable first approximation, though not necessarily a unique model, for the intrinsic Centaur distribution.

\section{Conclusions \label{Conclusions}}

We integrated the orbits of the 53~known Centaurs over 100~Myr, in a model that includes the Sun and the perturbations of the four giant planets self-consistently, using a symplectic integrator capable of handling close encounters with the planets.  The results shed light on the long term behavior of these objects, as well as their connection with Kuiper Belt objects~(KBOs), Scattered Disk objects~(SDOs), and Jupiter-family comets~(JFCs). Our conclusions are summarized as follows.

\begin{enumerate}
\item Our integrations suggest that two-thirds of these Centaurs will be ejected from the solar system (or will enter the Oort Cloud), while one-third will be injected into the JFC population;  a few percent are likely to impact one of the giant planets.  

\item These Centaurs do not diffuse into the dynamically stable region of orbital parameter space populated by resonant and classical KBOs, and make limited excursions into the region populated by SDOs.  This is not inconsistent with the hypothesis that the KBOs and/or the SDOs provide the source for the Centaur population, as objects may slowly ``leak'' out of the heavily-populated, relatively stable regions, but diffuse through the solar system before they are able to diffuse back in.  Based on the negligible diffusion of Centaurs to perihelion distance $q\gtrsim33$~AU, we suggest this as a dynamical division between Centaurs and SDOs.

\item The orbital evolution of this sample of Centaurs is strongly chaotic, and characterized by frequent close encounters with the planets. The process in which particles are ``handed off'' from the gravitational influence of one giant planet to another, as described by LD97, is less clearly seen in our results.  It is possible that this behavior does occur among Centaurs, but it may be less prominent in our study due to the higher inclinations and lower Tisserand parameters of our sample (compared to the initial conditions assumed in LD97). 

\item The known Centaurs do not exhibit long-term resonance sticking.  This is in contrast with the behaviour of SDOs.  Some Centaurs visit a number of different resonances but spend no more than a few~Myr in any single one, while others avoid resonances altogether.  This indicates that ``stable resonance islands'' take up a smaller fraction of Centaur phase space than is the case for SDOs.

\item The median dynamical lifetime of this sample is 9~Myr, but the individual lifetimes are highly variable; about 20\% of our sample have lifetimes shorter than 1~Myr, while another 20\% have lifetimes exceeding 100~Myr.  We found the average length of stay in the ``Chiron-type'' class \citep{Levison96} to be 6.5~Myr, while the same for the JFC class is 50,000~yr.

\item We find that these Centaurs spend most time at eccentricities between 0.2~and~0.6, and perihelia between 12~and 30~AU (Fig.~\ref{Contour}).  

\item Their time-weighted surface density (projected in the ecliptic plane) is nearly constant in the planetary region, and decreases beyond 30~AU approximately as a power law, $\sim r^{-2.5}$ (Fig.~\ref{Rdistrib}).

\item Characteristic inclinations (time-weighted values of $\cos^{-1}\langle\cos i\rangle$) are higher in some parts of parameter space than others, including some regions with high eccentricities and perihelia near Neptune, as well as in a fairly sizable region with semimajor axes near that of Uranus (Fig.~\ref{Idistrib}).  It is possible that high inclinations enhance the relative stability of the latter region by reducing the frequency of close encounters with the planets.

\item In most of the Centaur phase space (excluding only the highest eccentricities, $e \gtrsim 0.8$), the dynamics are nearly ergodic in our simulation.  Therefore, the time-weighted orbital distributions obtained from our simulation are not expected to be extremely sensitive to biases in initial conditions.  A simulated observational survey of the time-weighted model Centaur population yields orbital element distributions that are similar to those of the known Centaurs.  It indicates that orbits of high inclination and those of moderate and high eccentricity are likely under-represented in the known sample of Centaurs.  

\item The time-weighted distributions of our simulation (Figs.~\ref{Contour}, \ref{qi}--\ref{Icontour}) provide a possible -- not necessarily unique -- model for the intrinsic Centaur distribution.  A more comprehensive analysis of observational biases, and a larger set of initial conditions for dynamical models, would test the ergodic assumption further, and would improve estimates of the intrinsic Centaur distribution.

\end{enumerate}

\section{Acknowledgments}

We thank Hal Levison for providing the Swift-Skeel orbit integrator, and for his helpful review.  MST is supported by a National Science Foundation Graduate Research Fellowship.  RM acknowledges NASA grants NAG5-10346 and NAG5-11661.


\begin{thebibliography}{99}

\bibitem[Brown(2001)]{Brown} Brown, M.~E. 2001, AJ, 121, 2804

\bibitem[Dones, Levison, \&~Duncan(1996)]{Dones} Dones, L., Levison,~H.~F., \&~Duncan,~M. 1996, in ASP Conf. Ser.~107, \textit{Completing the Inventory of the Solar System}, ed.~T.~W.~Rettig \& J.~M.~Hahn (San Francisco: ASP), 233

\bibitem[Duncan, Quinn, \&~Tremaine(1987)]{Duncan87} Duncan, M., Quinn,~T., \& Tremaine,~S. 1987, AJ, 94, 1330

\bibitem[Duncan, Quinn, \&~Tremaine(1988)]{Duncan88} Duncan, M., Quinn,~T., \& Tremaine,~S. 1988, ApJ., 328, L69

\bibitem[Duncan, Levison, \&~Lee(1998)]{Duncan98} Duncan, M.~J., Levison,~H.~F., \& Lee,~M.~H. 1998, AJ, 116, 2067

\bibitem[Duncan \& Levison(1997)]{DL97} Duncan, M.~J., \& Levison,~H.~F. 1997, Science, 276, 1670

\bibitem[Fern\'andez(1997)]{Fernandez97} Fern\'andez, J.~A. 1997, Icarus, 129, 106

\bibitem[Fern\'andez, Gallardo, \&~Brunini(2002)]{Fernandez02} Fern\'andez, J.~A., Gallardo,~T., \& Brunini,~A. 2002, Icarus, 159, 358

\bibitem[Grazier \etal(1999)]{Grazier} Grazier, K.~R., Newman,~W.~I., Varadi,~F., Kaula,~W.~M., \& Hyman,~J.~M. 1999, Icarus, 140, 353

\bibitem[Gladman \etal(2001)]{Gladman} Gladman, B., Kavelaars,~J.~J., Petit,~J.-M., Morbidelli,~A., Holman,~M.~J., \& Loredo,~T. 2001, AJ, 122, 1051

\bibitem[Holman \& Wisdom(1993)]{Holman93} Holman, M., \& Wisdom,~J. 1993, AJ., 105, 1987

\bibitem[Kowal(1989)]{Kowal} Kowal, C. 1989, Icarus, 77, 118

\bibitem[Lecar \etal(2001)]{Lecar} Lecar,~M., Franklin,~F.~A., Holman,~M.~J., \& Murray,~N.~W. 2001, ARA\&A, 39, 581

\bibitem[Levison(1996)]{Levison96} Levison, H. 1996, in ASP Conf. Ser.~107, \textit{Completing the Inventory of the Solar System}, ed.~T.~W.~Rettig \& J.~M.~Hahn (San Francisco: ASP), 173

\bibitem[Levison \& Duncan(1993)]{LD93} Levison, H.~F., \& Duncan,~M.~J. 1993, ApJL, 406, L35

\bibitem[Levison \& Duncan(1997)]{LD97} Levison, H.~F., \& Duncan,~M.~J. 1997, Icarus, 127, 13

\bibitem[Malhotra(1996)]{Malhotra96} Malhotra, R. 1996, AJ, 111, 504

\bibitem[Malhotra, Duncan, \&~Levison(2000)]{Malhotra00} Malhotra, R., Duncan,~M.~J., \& Levison,~H.~F. 2000, in \textit{Protostars and Planets IV}, ed.~V.~Mannings, A.~P.~Boss, \& S.~S.~Russell (Tucson: Univ. Arizona Press), 1231.

\bibitem[Malyshkin \& Tremaine(1999)]{Malyshkin} Malyshkin, L., \& Tremaine,~S. 1999, Icarus, 141, 341

\bibitem[Millis \etal(2002)]{Millis} Millis, R.~L., Buie,~M.~W., Wasserman,~L.~H., Elliot,~J.~L., Kern,~S.~D., \& Wagner,~R.~M. 2002, AJ, 123, 2083

\bibitem[Morbidelli(1997)]{Morbidelli} Morbidelli, A. 1997, Icarus, 127, 1

\bibitem[Wiegert \& Tremaine(1999)]{Wiegert} Wiegert, P., \& Tremaine,~S. 1999, Icarus, 137, 84

\bibitem[Wisdom \& Holman(1991)]{Wisdom} Wisdom, J., \& Holman,~M. 1991, AJ, 102, 1528

\end{thebibliography}
\end{document}